\documentclass[iop,apj,twocolumn]{emulateapj_pab}	

\RequirePackage{hyperref}
\hypersetup{
	colorlinks = true,
	linkcolor = blue,
	citecolor = blue,
	filecolor = blue,
	urlcolor = blue,
	pdfborder = {0 0 0}
}

\usepackage{color,bm,mathrsfs,amsmath}

\renewcommand{\vec}[1]{\bm{#1}}

\graphicspath{{figures/}{../figures/}{MF_tripole/figures/}{MF_quadrupole/figures/}}

\submitted{Received 2018 April 11; revised 2018 October 12; accepted 2018 October 15; published 2018 December 5}

\journalinfo{The Astrophysical Journal, {\rm 869:3 (7pp), 2018 December 10 \hfill \url{https://doi.org/10.3847/1538-4357/aae97f}}}
\shorttitle{Magnetic Helicity from Multipolar Regions}
\shortauthors{Bourdin \& Brandenburg}

\newcommand{\tab}[1]{Table\,\ref{T:#1}}
\newcommand{\sect}[1]{Section\,\ref{S:#1}}

\DeclareRobustCommand*{\fig}[1]{Figure\,\ref{F:#1}}

\newcommand{\eqn}[1]{Equation\,(\ref{E:#1})}

\newcommand{\graphflex}[4][figure]{\begin{#1}#2\caption{#4\label{F:#3}}\end{#1}}
\newcommand{\graph}[3]{\begin{figure}\plotone{#1.pdf}\caption{#3\label{F:#2}}\end{figure}}
\newcommand{\graphwidthflex}[6][figure*]{\graphflex[#1]{#5\includegraphics[width=#4]{#2.pdf}}{#3}{#6}}
\newcommand{\graphwidth}[4][15cm]{\graphwidthflex{#2}{#3}{#1}{\centering}{#4}}
\newcommand{\graphfull}[3]{\graphwidth[17cm]{#1}{#2}{#3}}

\newcommand{\eql}[1]{\begin{equation}#1\end{equation}}
\newcommand{\eqa}[1]{\begin{eqnarray}#1\end{eqnarray}}

\newcommand{\eqi}[1]{$#1$}

\newcommand{\bra}[1]{\langle #1\rangle}
\newcommand{\braxy}[1]{\langle #1\rangle_{xy}}

\newcommand{\MMMM}{\mbox{\boldmath ${\sf M}$} {}}
\newcommand{\rrr}{\hat{\mbox{\boldmath $r$}} {}}
\newcommand{\eee}{\hat{\mbox{\boldmath $e$}} {}}
\newcommand{\hxxi}{\hat{\mbox{\boldmath $\xi$}}{}}{}
\newcommand{\heeta}{\hat{\mbox{\boldmath $\eta$}}{}}{}
\newcommand{\xxi}{\mbox{\boldmath $\xi$} {}}
\newcommand{\eeta}{\mbox{\boldmath $\eta$} {}}

\definecolor{darkgreen}{rgb}{0,0.45,0}

\newcommand{\upper}[1]{{\uparrow#1}}
\newcommand{\bound}[1]{{\downarrow#1}}

\begin{document}
\hypersetup{
	pdftitle = {Magnetic Helicity from Multipolar Regions on the Solar Surface},
	pdfauthor = {Philippe-A.~Bourdin},
	pdfkeywords = {Sun: magnetic fields -- Sun: corona -- dynamo -- magnetohydrodynamics (MHD) -- turbulence -- methods: numerical},
	pdfsubject = {The Astrophysical Journal, 869(2018) 3. doi:10.3847/1538-4357/aae97f}
}


\title{Magnetic Helicity from Multipolar Regions on the Solar Surface}

\author{Philippe-A.~Bourdin$^{1}$, \orcid{0000-0002-6793-601X},
Axel~Brandenburg$^{2,3,4,5}$, \orcid{0000-0002-7304-021X}
}
\email{Philippe.Bourdin@oeaw.ac.at}
\email{brandenb@nordita.org}

\affiliation{$^{1}$ Space Research Institute, Austrian Academy of Sciences, Schmiedlstr. 6, A-8042 Graz, Austria}
\affiliation{$^{2}$ Nordita, KTH Royal Institute of Technology and Stockholm University, Roslagstullsbacken 23, SE-10691 Stockholm, Sweden}
\affiliation{$^{3}$ JILA and Department of Astrophysical and Planetary Sciences, University of Colorado, Boulder, CO 80303, USA}
\affiliation{$^{4}$ Department of Astronomy, AlbaNova University Center, Stockholm University, SE-10691 Stockholm, Sweden}
\affiliation{$^{5}$ Laboratory for Atmospheric and Space Physics, University of Colorado, Boulder, CO 80303, USA}

\begin{abstract}
The emergence of dipolar magnetic features on the solar surface is
an idealization.
Most of the magnetic flux emergence occurs in complex multipolar regions.
Here, we show that the surface pattern of magnetic structures alone
can reveal the sign of the underlying magnetic helicity in the nearly
force-free coronal regions above.
The sign of the magnetic helicity can be predicted to good accuracy
by considering the three-dimensional position vectors of three spots
on the sphere
ordered by their relative strengths at the surface and compute from
them the skew product.
This product, which is a pseudoscalar, is shown to be a good proxy
for the sign of the coronal magnetic helicity.
\end{abstract}

\section{Introduction}

The Sun's magnetic field manifests itself through sunspots in
white light and magnetograms in polarized light.
The resulting pattern is generally rather complex and never in the form
of a symmetric pair of spots that is usually shown in text books;
see, e.g., \cite{Parker:1979}.
Just like a human face, both halves are never perfectly
identical to their actual mirror images.
In fact, certain aspects of the solar surface pattern are distinctly
different from each other in the northern and southern hemispheres.
This aspect has been explored by Sara \cite{Martin:1998a,Martin:1998b,Martin:2003}
and others \citep{Canfield+al:1999,Magara+Longcope:2001,Gibson+al:2002};
see also \cite{Panasenco+Martin:2008} and \cite{Panasenco+al:2011,Panasenco+al:2013} for recent accounts of those studies.

The purpose of this paper is to propose a simple recipe by which a
pseudoscalar can be constructed that can be used to estimate the sign
of the underlying magnetic helicity that is responsible for creating
the surface magnetic field.
To appreciate the motivation behind this way of thinking, we must recall
that a pseudoscalar is a rather special type of mathematical construct.
If one finds a way of constructing such a quantity from some physical
object, it changes its sign when constructing it from a mirror image of
the same object.
A common example is the cyclones on the Earth's weather map that look
different from their mirror images.
Mathematically, a pseudoscalar can be constructed from the downward
direction given by the gravity vector $\vec{g}$ and the local angular
velocity vector $\vec{\Omega}$, which is an axial vector.
If one draws this vector by indicating the sense of rotation rather than
through an arrow on one of the two ends, it is evident that viewing it
in a mirror results in the opposite sense of rotation.
Therefore, the sign of the dot product $\vec{g} \cdot \vec{\Omega}$ changes.
It is therefore a pseudoscalar.
In the Earth and in the Sun, it is this pseudoscalar that governs the
sign of several other relevant pseudoscalars such as the kinetic,
magnetic, and current helicities; see \cite{Brandenburg+Subramanian:2005} for a review.

One may now speculate that the construction of any pseudoscalar in
a system must be related to some other pseudoscalar in that system,
even though the causal relationship may not immediately be evident.
Consider now a seemingly absurd looking example of a pseudoscalar,
\eql{Q=(\hat{\vec{n}}_1\times\hat{\vec{n}}_2)\cdot\hat{\vec{n}}_3,\label{E:Qexpression}}
constructed from the normalized position vectors $\hat{\vec{n}}_1$, $\hat{\vec{n}}_2$, and $\hat{\vec{n}}_3$
of three different spots on the solar surface.
In this definition, \eqi{Q} is independent of the sphere radius.
Their unsigned fluxes are $|\Phi_1|$, $|\Phi_2|$, and $|\Phi_3|$, and
they are ordered such that $|\Phi_1|<|\Phi_2|<|\Phi_3|$.

Remarkably, the direction of the magnetic field does not enter, and we
do not even need a vector-magnetogram.
All that is required is any three spots that can somehow be ordered,
for example, by their strength, as explained above.

In this paper, we test the idea outlined above by considering synthetic
line-of-sight magnetograms and constructing from those a nearly force-free
magnetic field in the corona on top of it.
Such a magnetic field is in general always helical, so we can compute the
sign of the magnetic helicity and compare it with the sign of $Q$.
The idea of constructing a pseudoscalar $Q$ from the position vectors
of individual sources is not new and has been applied to the arrival
directions of energetic GeV photons coming from extragalactic sources in
the sky \citep[][who find evidence for a negative
sign throughout all of the sky, which they associated with the possibility
of a helical primordial magnetic field with negative helicity in all of
the Universe]{Tashiro+al:2014,Chen+al:2015,Tashiro+Vachaspati:2015}.
In their case, the arrival directions of GeV $\gamma$-rays in the sky
is the result of magnetic deflection of pair-created particles resulting
from the interaction of TeV photons from blazars with the extragalactic
background light.
In the present case, the location of spots on the solar or stellar
surface is a more direct consequence of a dynamo-generated magnetic
field somewhere beneath the surface \citep{Brandenburg:2005}.

In the solar corona, a force-free magnetic field can be constructed
using a potential field extrapolation method.
Only the line-of-sight magnetic field $B_z(x,y)$ at the bottom boundary
is needed.
No magnetic helicity can readily be constructed from this, and yet
a certain sign of the resulting magnetic helicity is somehow encoded in
the photospheric magnetic pattern, provided it looks different from its
mirror image, as alluded to above.
This is indeed what was found in the recent work of \cite{Bourdin+al:2018_helicity},
using data from simulations of \cite{Bourdin+al:2013_overview}.

Any distribution of $B_z(x,y,z_\ast,0)$ at the surface $z=z_\ast$ can be
used to construct a potential field $\vec{B}=\vec{\nabla}\varphi$, where
$\varphi(x,y,z)$ obeys $B_z(x,y,z_\ast,0)=\partial\varphi/\partial z$
at $z=z_\ast$.
At later times, however, $B_z(x,y,z_\ast,t)$ changes on the boundary
as the magnetic field patches expand due to diffusive and dynamical
processes.
The field then evolves through a sequence of new nonpotential
nearly force-free states, with the current density being parallel
to the magnetic field, which implies local current helicity.

The purpose of this work is to study the connection between the original
orientation of spots, as characterized by the pseudoscalar $Q$, and the
sign of the resulting magnetic helicity in the volume above.
For this purpose, we solve for the magnetic field numerically using
ambipolar diffusion as a relaxation method to construct approximately
force-free magnetic fields above three- and four-spot arrangements as
the lower boundary condition.

\section{Method}

\subsection{Ambipolar relaxation approach}

To construct an approximately force-free equilibrium magnetic field
$\vec{B}$, one often uses the magnetofrictional approach \citep{Yang+al:1986},
which corresponds to solving the induction equation with a velocity
that is proportional to the Lorentz force, $\vec{J}\times\vec{B}$, where
$\vec{J}=\vec{\nabla}\times\vec{B}/\mu_0$ is the current density and $\mu_0$ is the
vacuum permeability.
One usually divides this velocity by $\vec{B}^2$ to enhance the relaxation rate
in regions of weak magnetic field \citep{Valori+al:2007}, but this is
purposely ignored here,
so our effective advection velocity is $\vec{v}=(\tau/\rho)\,\vec{J}\times\vec{B}$,
where $\tau$ is some relaxation time and $\rho$ is some density, which is
assumed constant.
Inserting this into the induction equation, the electromotive
force $\vec{v}\times\vec{B}$ becomes proportional to $(\vec{J}\times\vec{B})\times\vec{B}
=(\vec{J}\cdot\vec{B})\,\vec{B}-\vec{B}^2\vec{J}$,
so the uncurled evolution equation for the magnetic vector potential
$\vec{A}$, where $\vec{B}=\vec{\nabla}\times\vec{A}$, can be written as
\eql{{\partial\vec{A}\over\partial t}=\alpha_{\rm AD}\vec{B}-(\eta_{\rm AD}+\eta)\mu_0\vec{J},
\label{E:dAdt}}
where $\alpha_{\rm AD}=(\tau/\rho)\,(\vec{J}\cdot\vec{B})$ is a term reminiscent
of the $\alpha$ effect in mean-field electrodynamics \citep{Krause+Rädler:1980},
$\eta_{\rm AD}=(\tau/\rho)\,\vec{B}^2$ is an effective magnetic diffusivity that
is what gives ambipolar diffusion its name, $\eta$ is the usual Spitzer
diffusivity, and the Weyl gauge has been adopted in \eqn{dAdt}.
Ambipolar diffusion is known to lead to the formation of sharp
structures such as current sheets between nearly force-free
regions in space \citep{Brandenburg+Zweibel:1994}.
This is an important feature that appears more pronounced here
than in the magnetofrictional approach, which motivates our
choice of employing ambipolar diffusion.

To formulate the potential field boundary condition, we employ
the Fourier-transformed magnetic vector potential
\eql{\tilde{\vec{A}}(k_x,k_y,z,t)=\int\vec{A}(x,y,z,t)\,e^{{\rm i}\vec{k}\cdot\vec{r}} {\rm d}^2\vec{r},}
where $\vec{k}=(k_x,k_y)$ and $\vec{r}=(x,y)$.
On the lower and upper $z$ boundaries, we thus have
\eql{{\partial\tilde{\vec{A}}\over\partial z}=-|\vec{k}|\tilde{\vec{A}}(k_x,k_y,z_\ast,t),}
where $z_\ast$ denotes the locations of the boundaries.
One of them is the lower surface, which we will from now on assume to
be at $z=0$, and the other is at the top of the domain.
Note that on both boundaries we assume the field to fall off with increasing
values of $z$, which is not the standard situation on the lower boundary,
if the region beneath it was supposed to be a vacuum.
It is, however, a natural choice in the present context where the
magnetic field is assumed to be initially potential inside the
computational domain \citep{Bourdin+al:2013_overview}.

We use this formulation to set the Fourier-transformed magnetic vector
potential in the ghost zones just outside the computational domain.
This corresponds to setting boundary conditions for the derivative of
all three components of $\tilde{\vec{A}}$, as stated above.
We solve \eqn{dAdt} numerically using the {\sc Pencil Code}\footnote{
\url{https://github.com/pencil-code}} using a resolution of $64^3$
meshpoints.
The total vertical magnetic flux is always zero since $\vec{A}$ is
periodic in $x$ and $y$, and we balance out $B_z$ at the bottom.

\subsection{Gauge-independent magnetic helicity}
To characterize the magnetic helicity in a gauge-independent fashion,
we use the formulation of
\cite{Finn+Antonsen:1985}; see their Equation,(5), which is identical to the
relative helicity of \cite{Berger+Field:1984}, and apply it to the
semi-infinite test volume \eqi{V(z)};
\eql{H_{\rm M}(z) = \iiint_{z}^{\infty}(\vec{A}+\vec{A}_{\rm pot}) \cdot (\vec{B}-\vec{B}_{\rm pot})\,{\rm d}z\,{\rm d}y\,{\rm d}x.}
We use a potential field extrapolation from the vertical
magnetic field \eqi{B_z} at the base of this test volume \eqi{V} as a
reference field, $\vec{B}_{\rm pot}=\vec{\nabla}\times\vec{A}_{\rm pot}$.
Computing the helicity difference between two test volumes above heights
\eqi{z} and \eqi{z+\Delta z}, we obtain the gauge-independent
magnetic helicity contained in small horizontal slices of thickness
\eqi{\Delta z} as
\eql{\Delta H_{\rm M}(z;\Delta z) = H_{\rm M}(z) - H_{\rm M}(z+\Delta z).}
Here, we simply use the grid distance of our simulation as
\eqi{\Delta z = 2 \pi / 64}.
We stop the integration at the upper boundary of our simulation domain, where
the magnetic field is almost potential, so that the error we make in our
limited integration is negligible.
As required by \cite{Finn+Antonsen:1985}, the magnetic fields normal to
the boundaries of the integration volume are identical.
Because our setup is periodic in the horizontal directions, all assumptions
on the boundaries of \eqi{V} apply only to the boundaries in \eqi{z}.

\subsection{Arrangement of spots of different strengths}

We consider configurations of $N$ spots of
different strengths $|\Phi_1|<|\Phi_2|<...<|\Phi_N|$ as
\eql{B_z(x,y,z_0,t)=\sum_{\lambda=1}^N\Phi_\lambda\,
\left. e^{-(\vec{r}-\vec{r}_\lambda)^2/2\sigma^2}\right/(2\pi\sigma_\lambda^2),}
where $\vec{r}_\lambda$ are the positions of the spots with magnetic fluxes
$\Phi_\lambda$ for $\lambda=1$, $2$, ..., $N$, where $N=3$ in this case.

We construct observables from spherical polar coordinates
$(r,\theta,\phi)$ at the surface as
\eql{\hat{\vec{n}}=(\sin\theta\cos\phi,\;\sin\theta\sin\phi,\;\cos\theta).}
To map the corners of a triangle onto the sphere, we choose
two neighboring unit vectors $\eee_1$ and $\eee_2$ on the sphere to define
a local coordinate system spanned by the unit vectors $\hxxi$ and
$\heeta$ given by $\xxi=\eee_2-\eee_1$ and $\eeta=\xxi\times\eee_1$.
Thus,
\eql{\hat{\vec{n}}_\lambda=\eee_1+\MMMM\rrr_\lambda
\quad\mbox{for $\lambda=1$, ..., $N$},}
where $\MMMM=(\hxxi,\heeta)$ is a $2\times3$ matrix consisting of
the two column vectors $\hxxi$ and $\heeta$.
In practice, we take $\eee_1=(1,0,0)$ and
$\eee_2=(\sin30\degr,\cos30\degr,0)$.
The matrix $\MMMM$ describes the conversion of planar two-dimensional
position vectors $\vec{r}_\lambda$ onto the position vector $\hat{\vec{n}}_\lambda$
on the unit sphere.
In this formulation, position differences $\vec{r}_\lambda-\vec{r}_{\lambda'}$
are then measured in radian.
To obtain the corresponding values in degrees, we multiply by
$180\degr/\pi$.

\subsection{Three-spot arrangements}
\label{S:Three-spot}

We begin with a triangular configuration with positions
\eql{\vec{r}_1=(\ell,0), \quad \vec{r}_2=(0,\ell), \quad \vec{r}_3=(\ell,\ell)
\quad\mbox{(A--C)},\label{E:triangle}}
with indices $(1,2,3)$ corresponding to spots of strengths
$\Phi_1/\Phi_0=1$, $\Phi_2/\Phi_0=2$, and $\Phi_3/\Phi_0=-3$,
where $\Phi_0$ is another constant and $\ell$ is the spot separation. 
This particular choice for the three values of $|\Phi_\lambda|$ ensures
vertical magnetic flux balance, although this is not actually required,
as will be shown further below.
We associate the positions $(\vec{r}_1,\vec{r}_2,\vec{r}_3)$ with
different permutations of the fluxes $(\Phi_1,\Phi_2,\Phi_3)$; see
\tab{cases}.

\begin{table}[t!]
\caption{Triangular cases with multiples of $\Phi_0$ at the positions $(r_1,r_2,r_3)$}
\vspace{-3pt}\centerline{
\begin{tabular}{lccrrrr}
\hline\hline
\# & $\vec{\Phi}$ & $Q/\ell^2$ & $H_{6\rm M+}$ & $H_{6\rm M-}$ & $H_{\rm C+}$ & $H_{\rm C-}$ \\
\hline
           A     &  $(3,-1,-2)$  & $+1$ & $\upper{+3.17}$ & $-2.22$          &  $+0.39$         & $-$\em{35.0}     \\
           A'    &  $(-1,3,-2)$  & $-1$ & $+2.22$         & $\upper{-3.17}$  &  $+$\em{35.0}    & $-0.39$          \\
${\mathscr A}$   &  $(3,1,2)$    & $+1$ & $\upper{+10.2}$ & $-0.23$          &  $+1.36$         & $-$\em{3.18}     \\
${\mathscr A}$'  &  $(1,3,2)$    & $-1$ & $+0.23$         & $\upper{-10.2}$  &  $+$\em{3.18}    & $-1.36$          \\
           B     &  $(-1,-2,3)$  & $+1$ & $+0.17$         & $-$\em{8.99}     &  $+0.07$         & $-$\em{10.9}     \\
           B'    &  $(-2,-1,3)$  & $-1$ & $+$\em{8.99}    & $-0.17$          &  $+$\em{10.9}    & $-0.07$          \\
${\mathscr B}$   &  $(1,2,3)$    & $+1$ & $+0.30$         & $-$\em{3.68}     &  $\bound{+4.46}$ & $-0.01$          \\
${\mathscr B}$'  &  $(2,1,3)$    & $-1$ & $+$\em{3.68}    & $-0.30$          &  $+0.01$         & $\bound{-4.46}$  \\
           C     &  $(-2,3,-1)$  & $+1$ & $+0.00$         & $-$\em{31.8}     &  $+8.78$         & $-$\em{15.0}     \\
           C'    &  $(3,-2,-1)$  & $-1$ & $+$\em{31.8}    & $-0.00$          &  $+$\em{15.0}    & $-8.78$          \\
${\mathscr C}$   &  $(2,3,1)$    & $+1$ & $+0.00$         & $-$\em{5.68}     &  $+0.70$         & $-$\em{3.88}     \\
${\mathscr C}$'  &  $(3,2,1)$    & $-1$ & $+$\em{5.68}    & $-0.00$          &  $+$\em{3.88}    & $-0.70$          \\
\hline
\tablenotetext{0}{\hspace*{-0.667em}Some numbers are displayed in italics to indicate a systematic trend.}
\label{T:cases}
\end{tabular}}\end{table}

The three spots, referred to as cases A--C in \eqn{triangle}, are arranged
in a mathematically positive (counterclockwise) sense around their center
of mass.
The same three-spot arrangement, but with only positive polarities,
will be referred to as cases ${\mathscr A}$--${\mathscr C}$.

In \tab{cases}, we compare the sign of $Q$ with the gauge-independent
magnetic helicity; left-handed systems ({\sf L}) have a positive $Q$
value and generate negative helicity signs.
We show the total magnetic (M) and current (C) helicities, where
we sum separately the positive ($+$) and negative ($-$) parts in our domain
that we compute as
\eqa{
H_{\rm M\pm} =& {\sum}_{z | \Delta H_{\rm M}(z) \gtrless 0} \Delta H_{\rm M}(z),& \nonumber \\
H_{\rm C\pm} =& {\sum}_{z | \Delta H_{\rm C}(z) \gtrless 0} \Delta H_{\rm C}(z),&}
with \eqi{\Delta H_{\rm C}(z)} being the total current helicity contained
in the horizontal slice of volume \eqi{4 \pi^2 \Delta z}.
In the tables, we give normalized values as
\eqi{H_{6\rm M\pm}\equiv10^{6}H_{\rm M\pm}}.
The total helicities of the whole domain may be obtained by summing the
negative and positive helicities:
\eqa{
H_{\rm M} = H_{\rm M+} + H_{\rm M-}\,, \nonumber \\
H_{\rm C} = H_{\rm C+} + H_{\rm C-}\,.}

Magnetic helicity values from the upper part of the domain are denoted by
$\uparrow$. The current helicity values denoted with $\downarrow$ are
strongly influenced by the lower boundary and are therefore disregarded.

It turns out that the value of $Q$ is proportional to $\ell^2$.
In particular, for the triangular arrangement of the spots given by
\eqn{triangle}, we have $Q=\pm\ell^2$ for positive (negative)
permutations of the spot's fluxes $(\Phi_1,\Phi_2,\Phi_3)$.
Conversely, for a flipped arrangement of spots (A'--C'),
we also find---not surprisingly---a flipped sign of $Q$.
These three spots are now arranged in a mathematically negative sense
around their center of mass.

Numerically, we find that for large values of $\ell$, corresponding
to angular separations of spots in excess of $20\degr$, the ratio
$|Q|/\ell^2$ drops significantly below unity.
In general, $Q/\ell^2$ is twice the area of the triangle spanned by the
three points, so, by comparison, for a configuration consisting of an
equilateral triangle, we have $Q/\ell^2=\pm\sin60\degr$ for spots of
increasing strength in the positive (negative) mathematical sense.

\graphfull{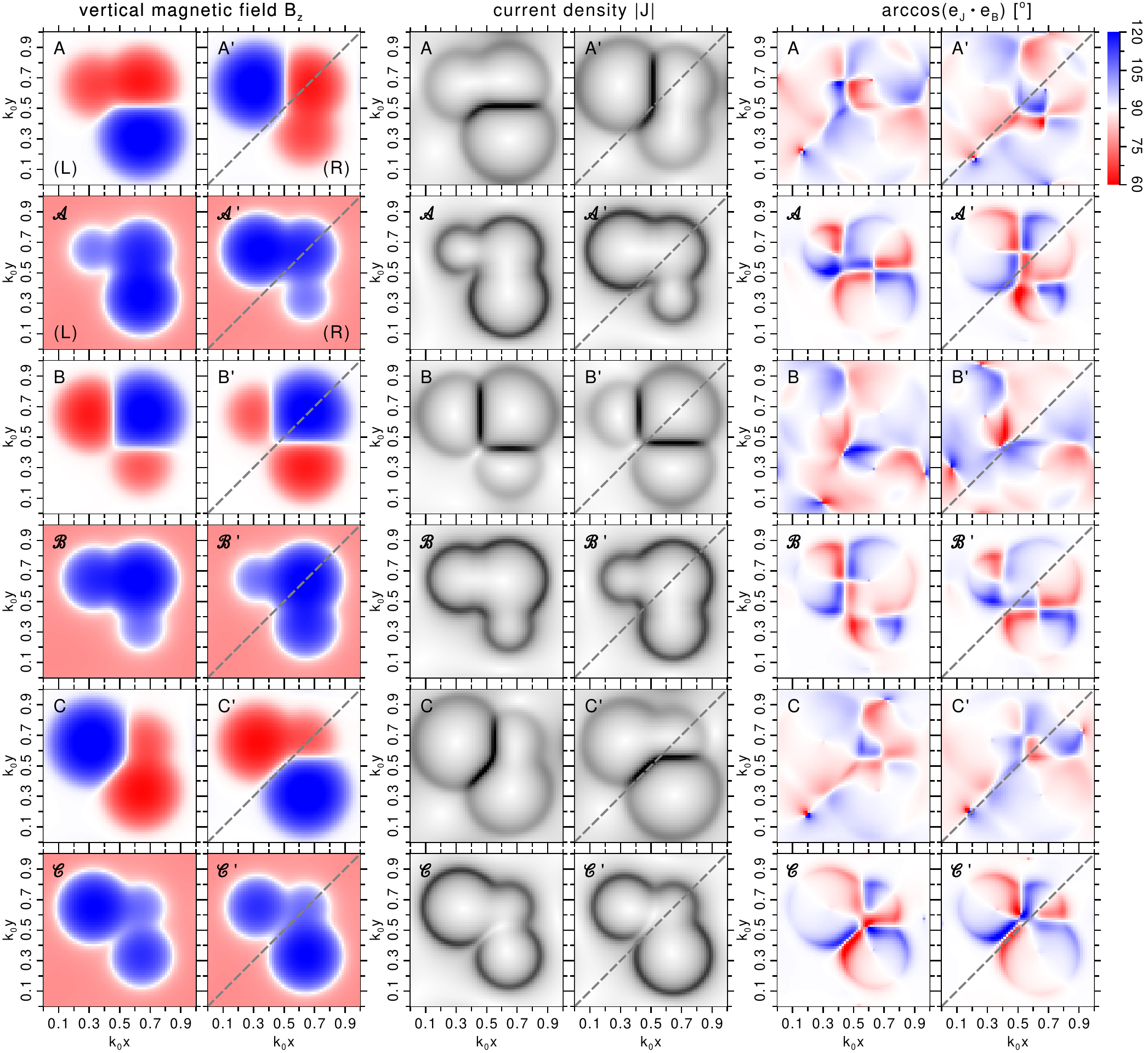}{p_overview}{Overview of $B_z(x,t)$ 
(red: $B_z<0$, blue: $B_z>0$), current
density (linear grayscale), and $\braxy{\bm{J}\cdot\bm{B}}$
at an evolved snapshot at $t=7$ time units at the bottom boundary.
The gray dashed line indicates the symmetry axis between cases
$X$ and $X'$; see also \tab{cases}.
The letters {\sf L} and {\sf R} in the panels of $|B_z|$ refer to
left and right hands, respectively. The handedness is {\sf L} in all
panels $X$ and {\sf R} for all $X'$; see \sect{application}.}

\graphwidth[16cm]{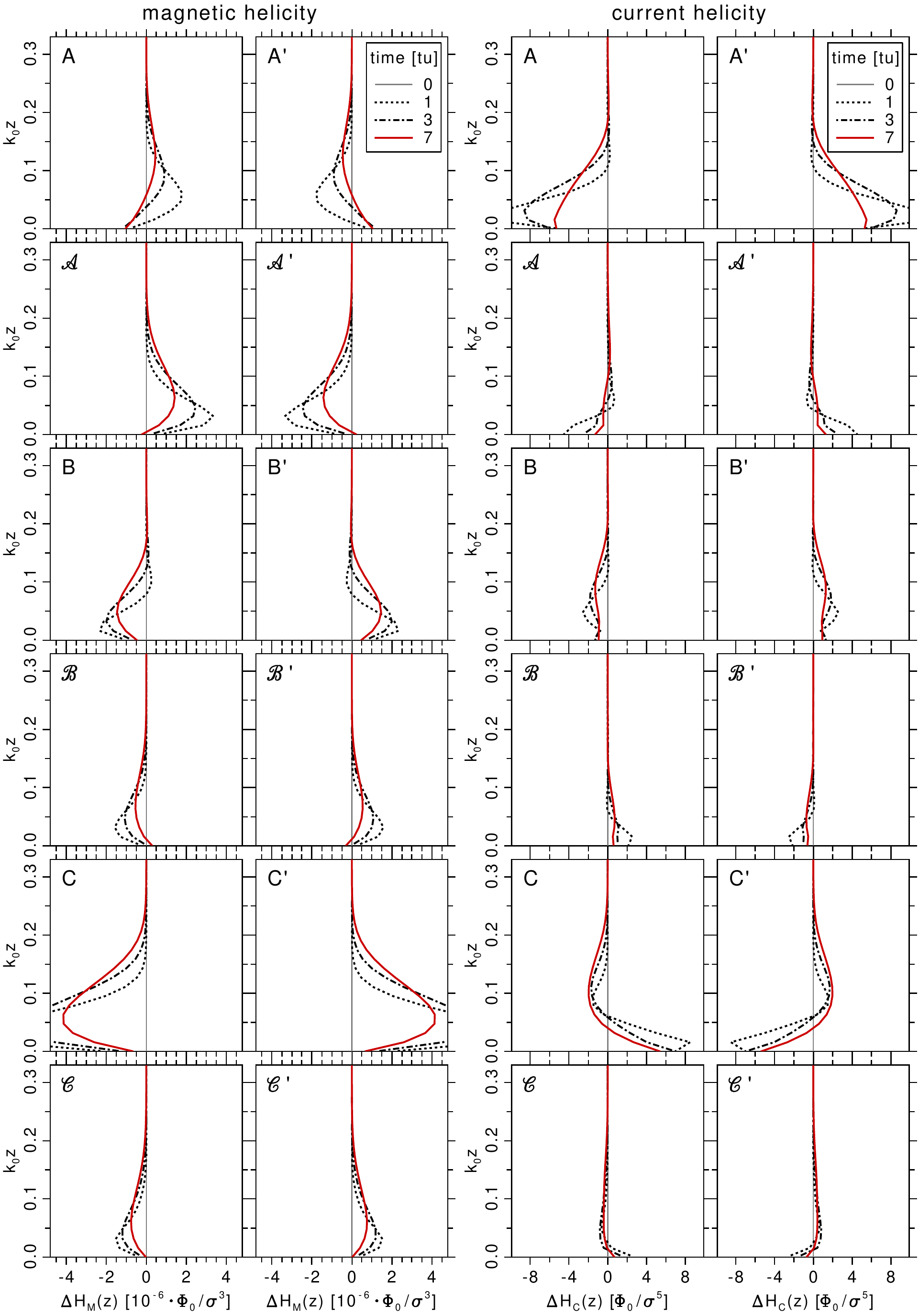}{p_antisym1}{
Overview of the magnetic and current helicity evolution for the signed
and unsigned three-spot configurations as listed in \tab{cases} at
three times. Time is given in arbitrary time units [tu].}

In \fig{p_overview}, we show visualizations of $B_z$, $|\bm{J}|$, and
$\bm{J}\cdot\bm{B}$ at the bottom surface.
We recall that the configurations A and A', B and B', as well
as C and C', each have opposite signs of current helicity.
We confirmed that the polarities of the different patches is not
important for the sign of the current helicity, but it is instead for the
orientation of spots of increasing strength.
In the visualizations of $B_z$, patches of increasing strengths
are arranged in a clockwise sense, while those of A'--C'
are in a counterclockwise sense.
Comparing A--C with A'--C', we see that the current sheets are flipped
about the diagonal, which is indicated by a long-dashed line.

The images of $\bm{J}\cdot\bm{B}$ look more complex than those of
$B_z$ and $|\bm{J}|$ because patches of opposite sign are now much
smaller than the patches of $B_z$.
However, one sees clearly that one is a mirror image of the other after
a sign flip of $\bm{J}\cdot\bm{B}$.
Apart from this, there is no change in the relative dominance
of one sign relative to the other, and one would not be able to tell the
sign of the $xy$ averages $\bra{\bm{J}\cdot\bm{B}}_{xy}$ judging just
based on the sign of the local patches of $\bm{J}\cdot\bm{B}$.
This demonstrates that what really matters in the end are both the sign
of $Q$ and the values of the unsigned flux of all spots, which is only
obtained after averaging over all the patches.

Remarkably, as alluded to above, the sign of the polarities is not
important for generating helicity.
In cases ${\mathscr A}$--${\mathscr C}$, where all spots have the same
polarity, we see that they also generate magnetic and current helicities,
but less strongly so than compared to the arrangements A--C that have
different signs in their polarities; see \fig{p_antisym1}.
The current densities of the unipolar cases ${\mathscr A}$--${\mathscr C}$
are of course stronger on the outer contour of the three-spot region.
The angles between $\bm{J}$ and $\bm{B}$ have a less complex pattern.
This is expected because of the less complex magnetic topology in the
unipolar regions.

\subsection{Four-spot arrangements}

In an attempt to generalize our approach to multiple spots,
we now also consider four-spot arrangements with
\eql{\vec{r}_1=(0,0), \; \vec{r}_2=(\ell,0), \; 
\vec{r}_3=(\ell,\ell), \; \vec{r}_4=(0,\ell), \;\label{E:quad}}
which are referred to as cases D--F'; see \tab{quad_cases}.
Each configuration is denoted by the vector
\eql{\vec{\Phi}=(\Phi_a,\,\Phi_b,\,\Phi_c,\,\Phi_d)}
for each of the positions $(\vec{r}_1,\vec{r}_2,\vec{r}_3,\vec{r}_4)$,
being certain permutations of $(\Phi_1,\Phi_2,\Phi_3,\Phi_4)$, where
the four fluxes obey $|\Phi_1|<|\Phi_2|<|\Phi_3|<|\Phi_4|$.

\begin{table}[t!]
\caption{
Quadratic cases with multiples of $\Phi_0$ at the positions $(r_1,r_2,r_3,r_4)$.}
\vspace{-3pt}\centerline{
\begin{tabular}{lclrrrr}
\hline\hline
\# & $\vec{\Phi}$ & $S$ & $H_{6\rm M+}$ & $H_{6\rm M-}$ & $H_{\rm C+}$ & $H_{\rm C-}$ \\
\hline
D   &  $(0.5,1.0,1.5,-3)$  & $+18$ & $+$\em{8.70}    & $-1.14$          &  $+$\em{31.0} & $-0.35$       \\  
E   &  $(0.5,1.0,-3,1.5)$  & $ -3$ & $+0.18$         & $-$\em{6.80}     &  $+0.08$      & $-$\em{12.9}  \\  
F   &  $(0.5,1.5,1.0,-3)$  & $ +1$ & $+1.10$         & $\upper{-3.01}$  &  $+$\em{14.4} & $-0.86$       \\  
E'  &  $(0.5,1.5,-3,1.0)$  & $ +3$ & $+$\em{6.80}    & $-0.18$          &  $+$\em{12.9} & $-0.08$       \\  
F'  &  $(0.5,-3,1.0,1.5)$  & $ -1$ & $\upper{+3.01}$ & $-1.10$          &  $+0.86$      & $-$\em{14.4}  \\  
\hline
\tablenotetext{0}{\hspace*{-0.667em}Some numbers are displayed in italics to indicate a systematic trend.}
\label{T:quad_cases}
\end{tabular}}\end{table}

A four-spot arrangement can be analyzed by breaking it down
into four different three-spot arrangements and calculating
the weighted sum
\eql{S=\sum_{i=1}^4 q_i \phi_i,}
where $q_i=Q_i/\ell^2=\pm1$ is the normalized $Q$ value for
triangle $i$ and $\phi_i$ is the total unsigned flux of each
triangle, i.e., \eql{\phi=\sum_{\lambda=a}^c |\Phi_\lambda|,}
which is different for each of the four triangles.
In \tab{quad_cases}, we compare the $S$ values for all possible
four-spot arrangements and find agreement with the signs and relative
magnitudes of the actual magnetic helicity in the lower part of the
domain.
Values denoted with $\uparrow$ are from the upper part of
the domain, where both signs contribute significantly in \fig{p_antisym2}.
We find that the sign changes of $S$ correlate well with those of the
generated helicities in the lower part of the domain.
Also, the magnitude of $S$ correlates with the strength of the generated
helicities.
When $S$ is small, the correlation with the helicities is less clear.
On the other hand, compared with the three-spot configurations,
the correspondence with the actual signs of current helicity,
and sometimes also the relative magnetic helicity, tends to be
the other way around.
The reason for this is not obvious, but the different trends for
magnetic and current helicities suggest that this is connected with
helicity effects from different length scales.
Therefore, investigating helicity spectra will be an important future task.

\graph{MF_quadrupole_overview}{p_antisym2}{Similar to \fig{p_antisym1},
but for the four-spot configuration listed in \tab{quad_cases}.}

\subsection{Application to solar magnetograms}
\label{S:application}

Nearly force-free magnetic field configurations have
been generated in numerical simulations by several groups
\citep{Gudiksen+Nordlund:2002,Gudiksen+Nordlund:2005a,Gudiksen+Nordlund:2005b,Bingert+Peter:2011,Bourdin+al:2013_overview}
using observed solar magnetograms as lower initial and boundary conditions,
where the vertical magnetic field was kept fixed.
They showed that random footpoint motions lead to field line braiding
and coronal heating.
Force-free magnetic fields are generally helical, but the initial
potential field and the random footpoint motions were nonhelical,
so no net sign of magnetic helicity was expected.
This turned out not to be the case.
Instead, these simulations produced net magnetic helicity, which
\cite{Bourdin+al:2018_helicity}
used to study the vertical variation of the resulting magnetic and current
helicity profiles and the possibility of a magnetic helicity reversal
with height, as has been suggested from studies of magnetic helicity in
the solar wind \citep{Brandenburg+al:2011} and theoretical studies
\citep{Warnecke+al:2011,Warnecke+al:2012}.
The point of the present work is to provide the theoretical underpinning
to why a finite value of the magnetic helicity can be expected in these
simulations that are otherwise statistically mirrorsymmetric.
In fact, the coronal simulations of \cite{Rempel:2017} also have finite
magnetic current helicity (M.\ Rempel 2018, private communication),
but in their case this could also be a remnant from the initial magnetic
field that was taken from a large-scale dynamo simulation.

The images of $B_z$ for ${\rm A}$ and ${\rm A}'$ in
\fig{p_overview} can be identified with those of open left ({\sf L})
or right ({\sf R}) hands, respectively.
Here, the palm corresponds to the biggest spot, the thumb points to the
smallest one, and the four fingers to the intermediate spot.
Their orientation indicates the expected sign of helicities, where
left-handed regions should generate negative helicity and right-handed
ones positive helicity.
The polarity of the spots is of no importance for the sign of the generated,
even though we find that mixed polarities generate more helicity.

Our work raises the possibility that magnetic helicity can be
determined even from a magnetic field with a $180^\circ$ ambiguity and, in
particular, from the relative arrangement of sunspots on the solar surface.
We have seen in \sect{Three-spot} that helicity is also generated if
all spots feature the same polarity, even though such setups create less
helicity than multipolar ones.
This suggests remarkable prospects for future work, whose full extent
cannot be imagined at present, given that detailed sunspot observations
exist for many centuries.

\section{Conclusions}

With the help of our model, we have demonstrated that magnetic helicity is generated above the
surface of a star like the Sun and that its sign can be determined uniquely from the horizontal
arrangement of magnetic flux concentrations and not just, as
previously thought, by the twist of the emerging magnetic field.
The source of this magnetic field is the solar dynamo, and it also
determines the sign of magnetic helicity.
This helicity manifests itself in a powerful way through the spot
arrangement.
We have demonstrated this by taking an observed solar magnetogram, removing all
other signs of magnetic helicity by fitting it to a potential field,
and then finding the original sign of magnetic helicity being recovered
through the spot arrangement alone \citep{Bourdin+al:2018_helicity}.
This illustrates a strong persistence of magnetic helicity
characterized by this new aspect of handedness.

An immediate implication of our work is that the horizontal arrangement
of just the line-of-sight magnetic field or---more precisely---the vertical
magnetic field, contributes to determining the sign of magnetic helicity
in the region above.
Such magnetograms have commonly been used in earlier studies of
coronal heating.
Although footpoint motions can lead to random twist of magnetic field
lines, the net effect vanishes, as has been confirmed in the simulations
of \cite{Bourdin+al:2013_overview}.
Nevertheless, net magnetic helicity has been detected in those simulations;
see \cite{Bourdin+al:2018_helicity}.
Our present results now give us a theoretical framework with which this
surprising fact can be understood.
In the simulations of \cite{Bourdin+al:2013_overview}, the underlying solar magnetogram
was taken from a location on the southern hemisphere slightly below the
solar equator.
Thus, if any net magnetic helicity is to be expected, it should be
positive \citep{Seehafer:1990}.
This is indeed what is found in the simulations.
This suggests that the sign of magnetic helicity must have been imprinted
on the pattern of the emerged magnetic flux at the surface through its
complex arrangement that cannot be modeled using only a pair of spots.
Three or more spots or magnetic flux concentrations are needed to encode
the information about magnetic helicity in the surface pattern.

Our work has future applications to solar physics and, perhaps, many
other fields.
Even just sunspot and starspot observations can in principle be used
to gather information about magnetic helicity.
Besides the original application of \cite{Tashiro+al:2014} to extragalactic
high-energy gamma rays, other possible applications in astrophysics may
include galactic magnetism, where magnetic helicity was previously only
accessible through Faraday rotation measurements \citep{Oppermann+al:2011,Brandenburg+Stepanov:2014}.
Other approaches are conceivable where the sign of magnetic helicity is in
principle accessible through the measurement of what is known as E and B
polarization \citep{Seljak+Zaldarriaga:1997,Kamionkowski+al:1997}, i.e., the parity-even and parity-odd
contributions to the linear polarization in the sky \citep{Kahniashvili+al:2014}.
The relation between this and our present work still needs to be
elucidated in more detail.


\acknowledgments
We thank the anonymous referee for useful suggestions and the
idea to associate the images of spot arrangements with the palm, fingers,
and thumb of left and right hands.
This work is financially supported by the Austrian Space Applications Programme at the Austrian Research Promotion Agency, FFG ASAP-12 SOPHIE under contract 853994.
This research was supported in part by the NSF Astronomy and Astrophysics
Grants Program (grant 1615100), and the University of Colorado through
its support of the George Ellery Hale visiting faculty appointment.
We acknowledge the allocation of computing resources provided by the
Swedish National Allocations Committee at the Center for Parallel
Computers at the Royal Institute of Technology in Stockholm.
This work utilized the Janus supercomputer, which is supported by the
National Science Foundation (award number CNS-0821794), the University
of Colorado Boulder, the University of Colorado Denver, and the National
Center for Atmospheric Research. The Janus supercomputer is operated by
the University of Colorado Boulder.

\bibliography{Literatur-PAB,literature-AB}
\bibliographystyle{aasjournal}

\end{document}